\documentclass[%
 reprint,
groupedaddress,
showpacs,preprintnumbers,
showkeys
nobibnotes,
 amsmath,amssymb,
 aps,
prb,
]{revtex4-1}

\usepackage{IEEEtrantools}
\usepackage{array}
\usepackage{amsmath,  amsfonts, amssymb}
\usepackage{graphicx}
\usepackage{dcolumn}
\usepackage{bm}
\usepackage{mathdots}
\usepackage{tikz}
\usepackage{overpic}

\begin{document}

\preprint{}

\title{Pair functions computed recursively in ordered and disordered lattices}

\author{T. Chattaraj}
 \altaffiliation[ ]
 \email{tirthaprasadchattaraj@gmail.com}

\noaffiliation

\date{\today}

\begin{abstract}
In this article I study pairing of two interacting particles in ideal 1D, 2D and Bethe lattices. I employ the method of recursion that has been formulated recently by Berciu et. al. \cite{berciu} to compute the pair functions in real space without performing any integrations. Although, in higher dimensions the system sizes that can be addressed with this method become limited, for disordered systems this size limit can be increased by employing approximations.
\begin{description}
\item[PACS numbers]
03.65.Ge, 31.15.aq, 71.10.Fd
\item[Keywords]
Recursion, Hubbard model, Bound pairs
\end{description}
\end{abstract}

\maketitle


\section{\label{intro}Introduction \protect\\  \lowercase{}}

Calculation of Green's functions of interacting particles is known to be a difficult task specifically in presence of impurities. Although these functions can be derived analytically using integral equations for one or two impurities \cite{economou}, for more impurities they are obtained much more easily through numerical procedures. The matrices involved for calculation of Green's functions of interacting particles in lattices with limited hopping range are generally sparse and there exists an efficient recursive method which uses this sparsity \cite{recursion}. This method based on continued fractions was applied as early as in the 1970s by Morita \cite{morita} and Haydock \cite{haydock} for calculation of density of states in ideal 3D lattices of cubic, fcc and bcc types. The recursion method has close resemblance in the formulation of Lanczos \cite{lanczos}. Recently, an efficient formulation has been developed by Berciu et. al. \cite{berciu} which I apply to disordered systems and graph structures as Bethe lattices. I present few calculations of bound pair spectral weights for the interacting Hubbard model on 1D, 2D and Bethe lattices using the method. 

Both attractive and repulsive interactions are known to be responsible for binding single particles into pairs \cite{cooper} \cite{winkler}. These pairs show different correlations depending on their statistics and strength of interaction \cite{greiner} \cite{bloch} \cite{lahini1} \cite{tirtha1}. The effect of this pairing leads to emergent phenomena as superconductivity \cite{cooper} \cite{anderson}. The most simplified model to understand this pairing and dependence of pairing on interaction strength can be obtained within the picture of two interacting particles. The pair functions describing two-particle spectral weights are obtained within a range of bandwidth for 1D and 2D Hubbard model of single level systems. The signature spectral weights  indicate the most important physics of full many particle systems \cite{rausch} within most simplified picture. 

The effect of pairs on the transport properties in regular lattice systems can be extended to general graph structures as Bethe lattice. The Bethe lattice, though interesting for it's mathematical form, similar forms are found at the cores of several photosynthetic complexes where the excitation transport is known to have effects of quantum interference \cite{engel} \cite{sarovar}. Thus study of quantum transport on the Bethe lattice at different interaction strengths between excitations may hold relevance to biological systems. Frequency dependence control of the quantum transport  on these structures can lead to the way for novel application using general graph architectures. 

 In disordered systems, the bound pairs are known to exhibit different behaviour of transport properties at different interaction regimes. In 1D lattices, an enhancement of delocalization at weak interaction regime and enhancement of localization at strong interaction regime has been noted by several calculations \cite{schreiber} \cite{pichard} \cite{flach} \cite{tirtha2}. In higher dimensions these calculations become difficult to perform and approximations are generally invoked \cite{ortuno}. The recursion method can be used to compute pair functions at any interaction strength for sufficiently large lattice systems in higher dimensions. The presence of external magnetic field can also be accounted directly with the method as described by Berciu et al. \cite{berciu}. It is exact and more efficient when compared with full diagonalization and application of approximations can make the calculations even more efficient within acceptable range of accuracy. The method is briefly described on next section and then few pair functions  of ideal 1D, 2D and Bethe lattices are presented  followed by applicability of approximations on disordered systems.

\section{The recursion algorithm}

A Green's function for a Hamiltonian $H$ is defined as
     \begin{equation}\label{gf}
              G(\omega) = \frac{1}{\omega - H},
       \end{equation} 
    where $\omega = E + \imath\eta$ is a complex number with $\eta$ a very small positive real number while the two particle Green's function $ G_2(m,n,\omega) = \langle m n | G(\omega) | m' n' \rangle $
is the propagator for two particles from sites $m'$, $n'$ to sites $m$, $n$ in a lattice, where $| m n \rangle = c_m^\dagger c_n^\dagger |0\rangle$ and the initial site indices $m'$, $n'$ are omitted for brevity. 
 For a Hubbard Hamiltonian
  \begin{equation}\label{ham}
H = \sum_{m} \epsilon_{m} a_{m}^\dagger a_m  + \sum_{<mn>} t_{|m-n|} a_{m}^\dagger a_n + \sum_{<mn>} v_{|m-n|} a_{m}^\dagger a_{n}^\dagger a_n a_m,
 \end{equation} 
 with nearest neighbour hopping and interaction, the recurrence relations connect the functions like $G_2(m, n, \omega)$ to functions $G_2(m-1, n, \omega), G_2(m+1, n, \omega), G_2(m, n-1, \omega)$ and $G_2(m, n+1, \omega)$. The bound pair spectral weight can be calculated for any local state from the Green's element $G_2(m', n', \omega)$ in real space  with $|m' - n'|=\{1,0\}$. 
   \begin{equation}\label{spec}
     A_2 (m', n', \omega) = \frac{-1}{\pi}  \mbox{Im}[G_2(m', n', \omega + \imath\eta)]
    \end{equation}
The density of states (DOS) can also be calculated from the spectral weights on real space.
   \begin{equation}\label{dos}
     DOS (\omega)  = \frac{1}{2}\sum_{m', n'} A_2 (m', n', \omega)
    \end{equation}
When there is  translational symmetry present in the system, only few local states with increasing relative distance $(|m' - n'|)$ might be sufficient for converging results. The vectors with a constant $m+n = R$ can be grouped within a vector which forms a 1D chain of $R$ and the Hamiltonian generates the recursion equation
   \begin{equation}\label{vrec}
    \mathbf{V}_{R} = \boldsymbol{\alpha}_R \mathbf{V}_{R-1} + \boldsymbol{\beta}_R \mathbf{V}_{R+1} + \mathbf{C},
    \end{equation}
with the hopping matrices $\boldsymbol{\alpha}_R$, $\boldsymbol{\beta}_R$ containing the hopping terms between the Green's elements. Here $\bf C = 0$ ( or $\neq \bf 0$) when $R\neq$ $m' + n'$ (or $= m' + n' = R'$). Solving this equation one can obtain all the Green's elements. At the boundaries of the chain
   \begin{equation}\label{vlr}
     \mathbf{V}_{0} =  \boldsymbol{\beta}_0 \mathbf{V}_{1}   \mbox{     and     } \mathbf{V}_{L} = \boldsymbol{\alpha}_L \mathbf{V}_{L-1},
    \end{equation}
where $0$ and $L$ are the minimum and maximum index possible for $R$. Simplifying Eq. \ref{vrec} as in Eq. \ref{vlr} with
   \begin{equation}\label{vr}
    \mathbf{V}_{R} = \mathcal{A}_R \mathbf{V}_{R-1}   \mbox{   and   }  \mathbf{V}_{R} = \mathcal{B}_R \mathbf{V}_{R+1},   \mbox{             for  } R \neq R',
    \end{equation}
the calculations become a recursion of vectors with
   \begin{equation}\label{rl}
 \mathcal{B}_{R} =  \left[1 -  \boldsymbol{\alpha}_R \mathcal{B}_{R-1}\right]^{-1} \boldsymbol{\beta}_R 
\mbox{,     } 
  \mathcal{A}_{R} =  \left[1 -  \boldsymbol{\beta}_R \mathcal{A}_{R+1}\right]^{-1} \boldsymbol{\alpha}_R .
    \end{equation}
These $\mathcal{A}_{R}$ and $\mathcal{B}_{R}$ matrices can be computed recursively starting from Eq. \ref{vlr} before one reaches $R=R'$ from both sides of the chain with  
   \begin{eqnarray}\label{vr'}
  \mathbf{V}_{R'} = \left[1 - \boldsymbol{\alpha}_{R'}\mathcal{B}_{R'-1} - \boldsymbol{\beta}_{R'}\mathcal{A}_{R'+1}\right]^{-1} \mathbf{C}.
    \end{eqnarray}
at $R= R'$. Once $\mathbf{V}_{R'}$ is found, all other  $\mathbf{V}_{R}$ can be found by Eq. \ref{vr}, hence solving the problem of finding all the Green's elements for a given $m'$ and $n'$ for a single $\omega$. The procedure accounts for the full self energy term which can be obtained from renormalized perturbation expansion on any ordered or disordered lattice. Computation of single particle Green's functions on any ordered or disordered lattices can also be performed with this procedure. For $H$ with long range hopping \cite{moeller}, Eq. \ref{vrec} can be extended to include the vectors beyond nearest neighbours.
   \begin{eqnarray}\label{lrvrec}
     \mathbf{V}_{R} &= \boldsymbol{\alpha}_{R,R-p} \mathbf{V}_{R-p} + \hdots +  \boldsymbol{\alpha}_{R,R-1} \mathbf{V}_{R-1}\nonumber\\
 & + \boldsymbol{\beta}_{R,R+1} \mathbf{V}_{R+1} + \hdots + \boldsymbol{\beta}_{R,R+p} \mathbf{V}_{R+p} + \mathbf{C}.
    \end{eqnarray}
 The recursion can be performed with the group of these vectors 
 \begin{equation}\label{lrgv}
    \Tilde{\mathbf{V}}_{R} = \left( \mathbf{V}_R, \cdots, \mathbf{V}_{R,R+p-1} \right).
    \end{equation}
   \begin{equation}\label{lrrec}
     \mathcal{Z}_R \Tilde{\mathbf{V}}_{R} = \mathcal{L}_R \Tilde{\mathbf{V}}_{R-p} + \mathcal{R}_R \Tilde{\mathbf{V}}_{R+p} + \mathbf{C}
    \end{equation}
with $\mathcal{Z}_{R}, \mathcal{L}_{R}$ and $\mathcal{R}_{R}$ containing several of the hopping matrices. This grouping of Green's functions can also be applied for the cases where sites have internal structures.

\begin{widetext}

\begin{figure}[h]
\includegraphics[width = 0.335\textwidth]{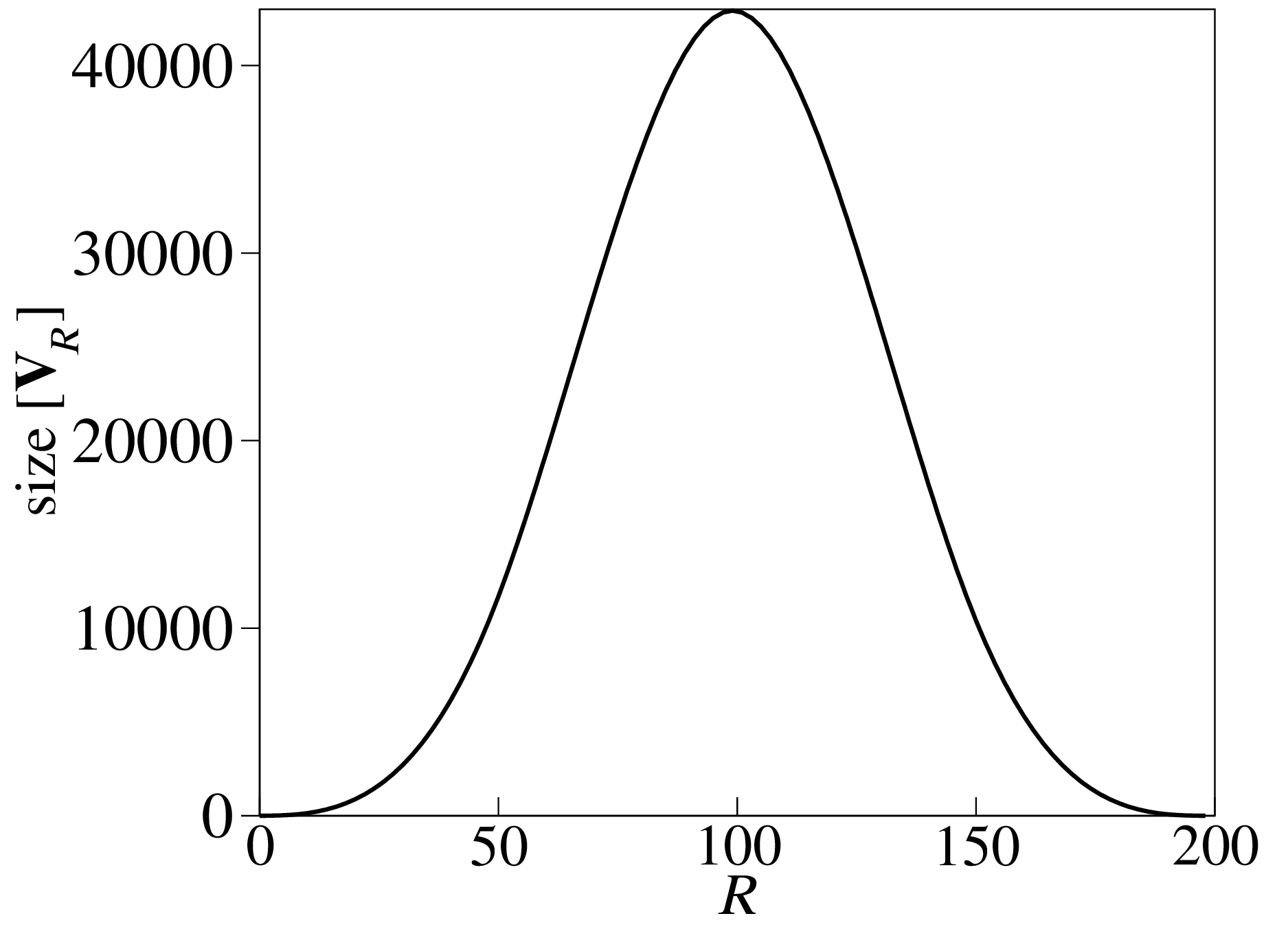}
\includegraphics[width = 0.325\textwidth]{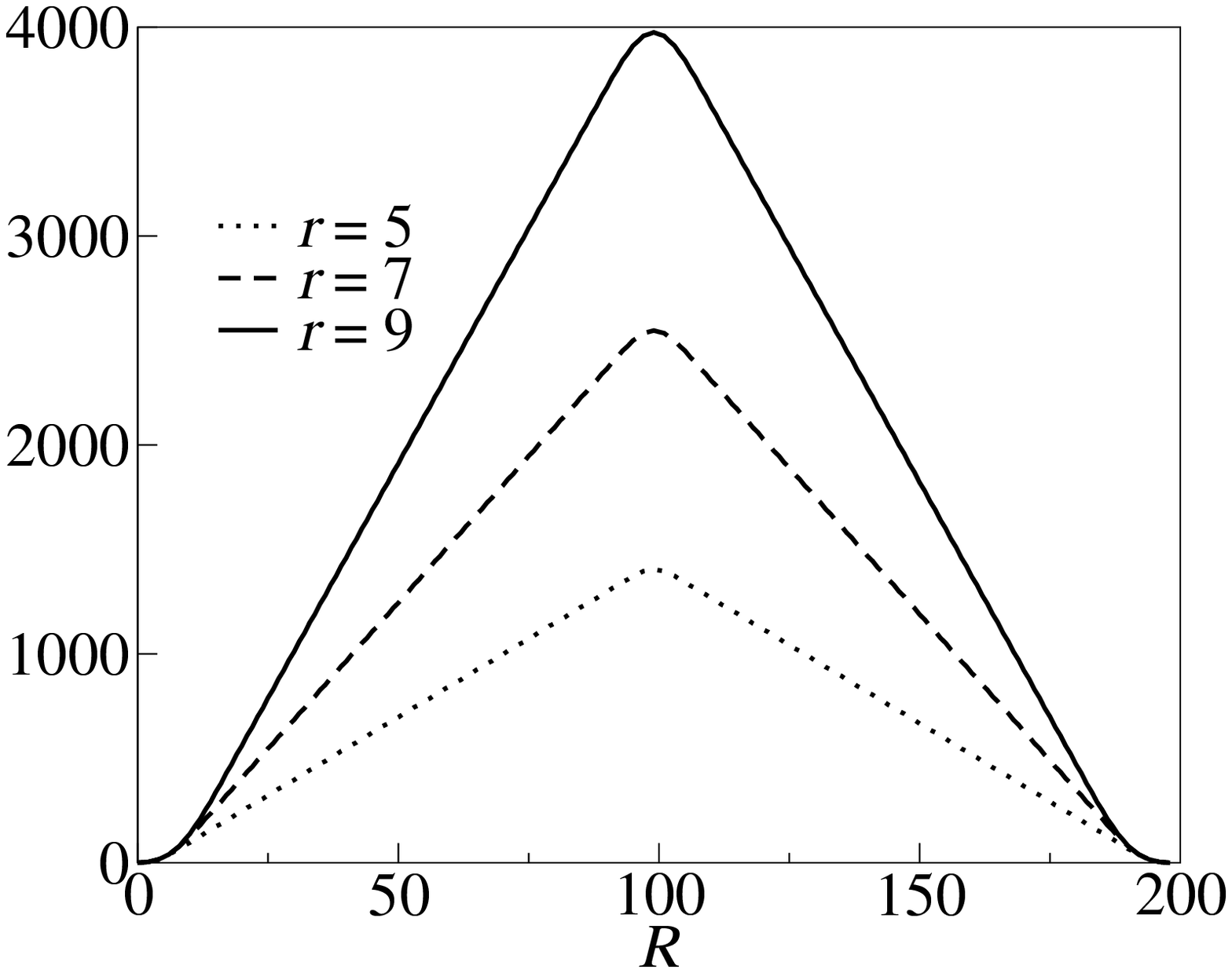}
\includegraphics[width = 0.325\textwidth]{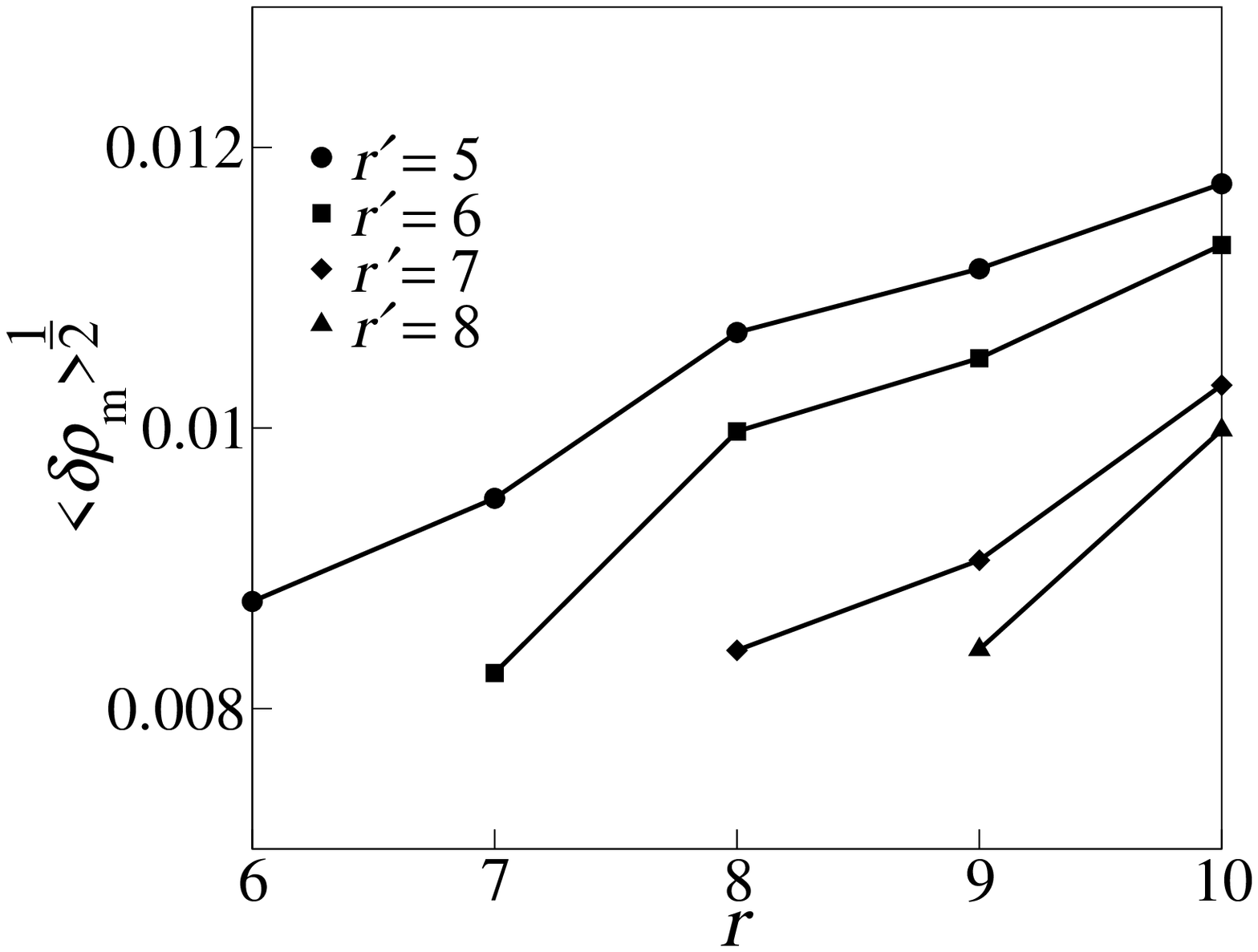}
\caption{Left panel - size of the vectors in recursion for a 2D lattice of 50 sites per dimension. Middle panel - size of the vectors with approximations. Right panel - Average absolute difference of the density between different maximum relative distances.}
\label{2D-disorder}
\end{figure}

\end{widetext}

Although, this recursion method allows for computation of pair functions for large lattice systems in low-dimensions, for higher dimensions, the system sizes that can be considered stays limited. These system sizes for ideal systems are generally large enough to eliminate significant errors due to finite size effects. In disordered systems, one often is interested in averaged properties \cite{tirtha2} of the Green's functions. For such computations, a limited system size becomes a hindrance in obtaining accurate results which can be surmounted with use of approximations. A physical approximation can be applied learning from previous works  \cite{lahini2} \cite{tirtha1} on pair functions where disorder was found to enhance correlations between particles. This translates to limiting relative distances between each particles  of the pair while computing the properties of disordered systems. This approximation allows for removing pair functions which have inter-particle distances greater than a step distance $r$.
\begin{equation}
G(n,m; n',m'; \omega)  \simeq 0     ~~~~~\mbox{ for }~~~~~ |n-m| > r
\end{equation}
With this approximation, the difficulty of computation of Green's function for a single point of $\omega$ scales as $\mathcal{O} (r^3N)$ for 1D systems and as $\mathcal{O} (r^3N^4)$ for 2D systems (for $r \ll N$). Thus gaining a factor of as $N^5$ in 1D lattices and as $N^8$ in 2D lattices compared to full diagonalization while producing accurate results. This makes the recursive method a preferable choice for calculation of Green's functions of large disordered systems in high dimensions. Accounting only the correlated Green's elements with a maximum relative distance $r$ between the particles doesn't attribute significant errors with the computed density distribution on the disordered lattice. The density terms are obtained within single particle sector projected from pair probabilities. The case for a 2D disordered lattice with disorder strength $W = 1$ and various $r$ is shown in Fig. \ref{2D-disorder}. The average of density difference per site is very small with changes of $r$ while the reduction of the size of the vectors and hence computational difficulty is significant with each unit decrease in $r$. 

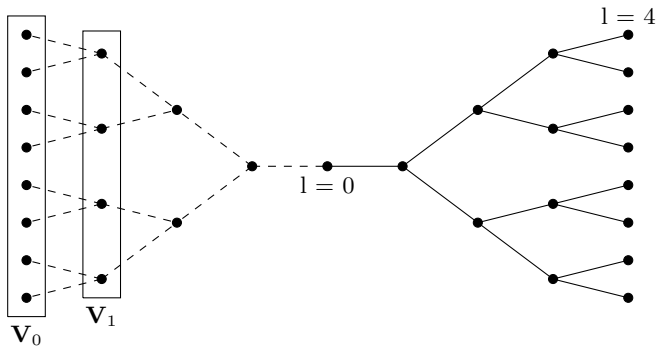
\begin{figure}[h]
\centering
\hspace{0.0cm}
\begin{tikzpicture}[scale=1.0]
\draw [fill] (0,0) circle [radius=0.06]; \draw [fill] (0,0.5) circle [radius=0.06]; \draw [fill] (0,1) circle [radius=0.06]; \draw [fill] (0,1.5) circle [radius=0.06]; \draw [fill] (0,2) circle [radius=0.06]; \draw [fill] (0,2.5) circle [radius=0.06];
\draw [fill] (0,3) circle [radius=0.06]; \draw [fill] (0,3.5) circle [radius=0.06]; \draw [fill] (1,0.25) circle [radius=0.06];
\draw [fill] (1,1.25) circle [radius=0.06]; \draw [fill] (1,2.25) circle [radius=0.06]; \draw [fill] (1,3.25) circle [radius=0.06]; \draw [fill] (2,1) circle [radius=0.06]; \draw [fill] (2,2.5) circle [radius=0.06]; \draw [fill] (3,1.75) circle [radius=0.06]; \draw [fill] (4,1.75) circle [radius=0.06]; \draw [fill] (5,1.75) circle [radius=0.06]; \draw [fill] (6,1) circle [radius=0.06]; \draw [fill] (6,2.5) circle [radius=0.06]; \draw [fill] (7,0.25) circle [radius=0.06]; \draw [fill] (7,1.25) circle [radius=0.06]; \draw [fill] (7,2.25) circle [radius=0.06]; \draw [fill] (7,3.25) circle [radius=0.06]; \draw [fill] (8,0) circle [radius=0.06]; \draw [fill] (8,0.5) circle [radius=0.06]; \draw [fill] (8,1) circle [radius=0.06];
\draw [fill] (8,1.5) circle [radius=0.06]; \draw [fill] (8,2) circle [radius=0.06]; \draw [fill] (8,2.5) circle [radius=0.06];
\draw [fill] (8,3) circle [radius=0.06]; \draw [fill] (8,3.5) circle [radius=0.06]; 
\draw (4,1.75) --(5,1.75); \draw (6,1)  --(5,1.75); \draw (6,2.5)  --(5,1.75); \draw (6,1)  --(7,0.25); \draw (6,1)  --(7,1.25); \draw (6,2.5)  --(7,2.25); \draw (6,2.5)  --(7,3.25); \draw (8,0)  --(7,0.25); \draw (8,1)  --(7,1.25); \draw (8,2)  --(7,2.25) ; \draw (8,3)  --(7,3.25) ; \draw (8,0.5)  --(7,0.25); \draw (8,1.5)  --(7,1.25); \draw (8,2.5)  --(7,2.25); \draw (8,3.5)  --(7,3.25); \draw [dashed] (3,1.75) --(4,1.75); \draw [dashed] (3,1.75) --(2,2.5); \draw [dashed] (3,1.75) --(2,1); \draw [dashed] (1,2.25) --(2,2.5); \draw [dashed] (1,0.25) --(2,1); \draw [dashed] (1,3.25) --(2,2.5); \draw [dashed] (1,1.25) --(2,1); \draw [dashed] (1,2.25) --(0,2); \draw [dashed] (1,0.25) --(0,0); \draw [dashed] (1,3.25) --(0,3); \draw [dashed] (1,1.25) --(0,1); \draw [dashed] (1,2.25) --(0,2.5); \draw [dashed] (1,0.25) --(0,0.5); \draw [dashed] (1,3.25) --(0,3.5); \draw [dashed] (1,1.25) --(0,1.5); 
\draw (0.75,0) -- (1.25,0) -- (1.25,3.55) -- (0.75,3.55) -- (0.75,0); 
\draw (-0.25,-0.25) -- (0.25,-0.25) -- (0.25,3.75) -- (-0.25,3.75) -- (-0.25,-0.25); 
\node at (4,1.5) {l = 0}; \node at (8,3.75) {l = 4}; \node at (1,-0.25) {$\mathbf{V}_1$}; \node at (0,-0.5) {$\mathbf{V}_0$};
\end{tikzpicture}
\vspace{0.5cm}
\caption{A Bethe lattice split into the left (dashed lines) and right (full lines) branches from root level $l=0$. The elements of each level of any branch can be combined into a vector which becomes a node in the chain of recursion.}
\label{Bethe}
\end{figure}

The recursion formulation maps directly for generalized application to Bethe lattices and similar structures as shown in Fig. \ref{Bethe}. The scheme for long-range hopping model would be applicable when intra-vector and nearest neighbor inter-vector Green's elements of these structures are coupled. The formulation has only three necessary components - two boundaries, an initial element and a chain of coupled vectors. This requirements are generally common with various calculations of lattices in low and high dimensions. For Bethe lattices this can be achieved with branching the lattice from root level ($l=0$) to the boundaries which forms both directions of a chain of vectors. For pair of particles, the sum of the indices of vectors form the indices for reformed vectors in chain.

\section{Spectral weights of bound pairs}
In this article I present several calculations on pair functions for the bound state. The case of ideal 1D, 2D and Bethe lattices are presented in Fig. \ref{ideal}. The left panel of Fig. \ref{ideal} explains the spectra for bound pairs in 1D ideal lattices for a single band Hubbard model with $\epsilon=0$ and $t=1$. For systems with very large on-site interaction ($|v_0| \gg 1$) and comparable nearest neighbour interaction with hopping strengths ( i. e. $|v_1| \approx 1$), the bound pairs describing two particles on same site will appear far apart from continuum states while the pairs bound at nearest neighbour will appear within continuum. These bound pairs within continuum and outside continuum have significantly different character. While a gap in the spectra for $|v_0| \gg 1$ induce Mott insulator phase, the bound pairs within continuum for $|v_1| \approx 1$ can be responsible with enhanced conductivity. The tunneling of pairs through single impurity is enhanced at weak interaction compared to non-interacting cases \cite{tirtha1}. The lifetime of bound pairs are known to be very high \cite{strohmaier} and thus such pairs can be responsible for conductance with less dissipation. The specific cases of $v_1=0$ and $v_1=2$ also shows significant differences in spectra. While the $v_1=0$ case is proportionate with full density of states, at $v_1=2$ the bound pairs are peaked at $\omega =2$ with a fast decay at $\omega > 2$. This characteristic peaking at $\omega \approx v_1$ with higher proportion of bound pairs generally leads to the hypothesis that to maximize the weight of bound pair states, weight of the states from Fermi level at $\omega =0$ to $\omega = v_1$ can be removed from the system for $v_1 <0$ and by increasing weight of the states to $\omega \ge v_1$ for $v/t >0$. For systems with negligible nearest neighbour interaction $v_1 \ll 0$ and comparable onsite interaction $v_0 \approx t$, the same physics will follow except $v_0 > 4$ will describe onset of Mott insulator phase. This simple description of pairs (which are boson like) points away from the necessity of the condensation criteria for enhancement of conductivity. For maximized pair states, the proportion of pair states compared to normal states can be calculated when particles are removed to $\omega \approx v_1$ from Fermi level.  For 2D ideal lattices the bandwidth is double that of 1D models and the bound pair peaks are broader which appear around $v_1$ and split from continuum for $v_1 \ge 8$. In the Bethe lattices the bound pair spectra show discontinuity at weak interaction regimes at $v_0 \approx 0$ and $v_1 \approx 0$. The $\omega$ dependence of bound pair proportion is very different from 1D and 2D continuous spectra. However at strong interactions these spectral lines merge to produce a single profile. 

\begin{widetext}

\begin{figure}[h] 
\includegraphics[width = 0.328\textwidth]{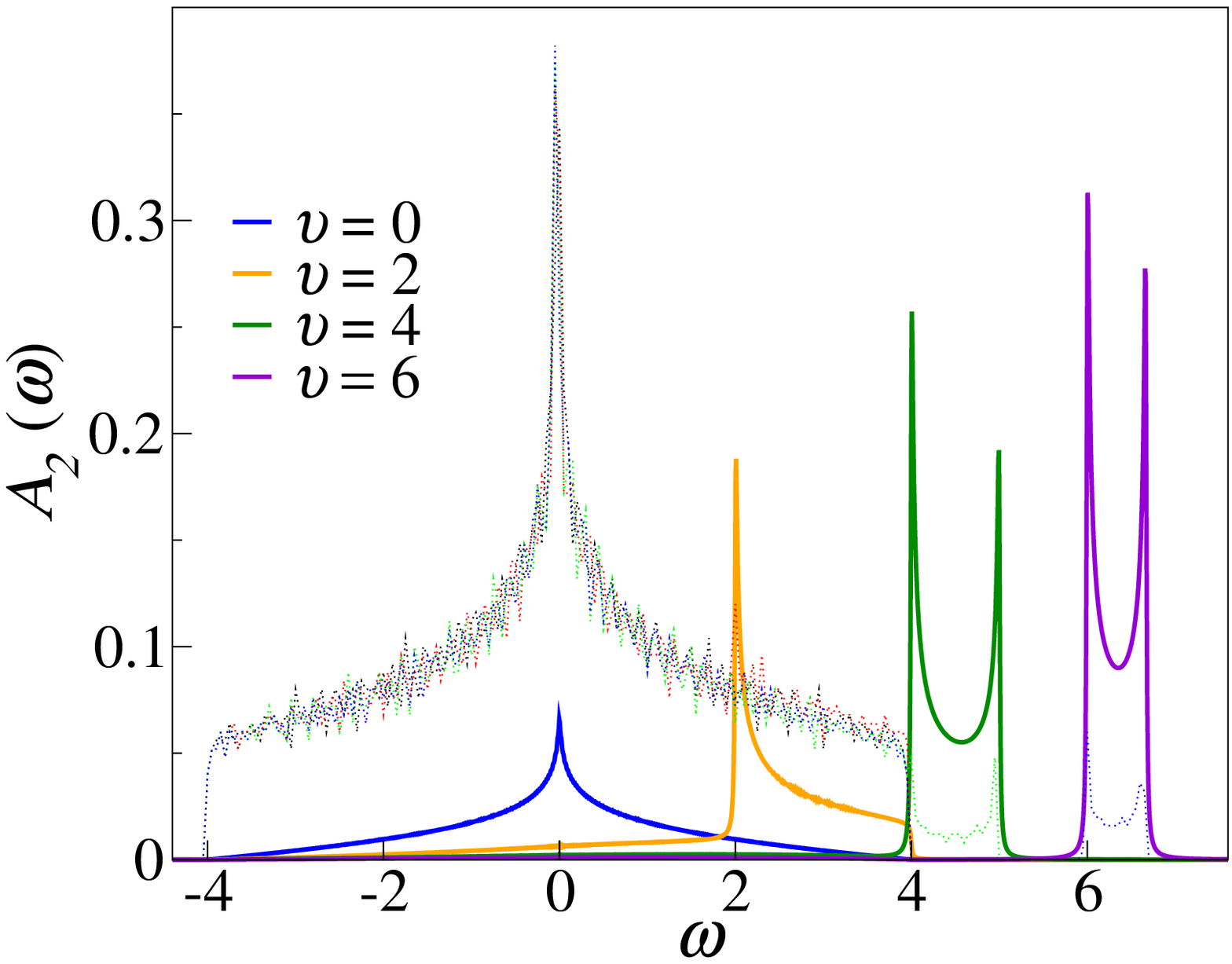}
\includegraphics[width = 0.328\textwidth]{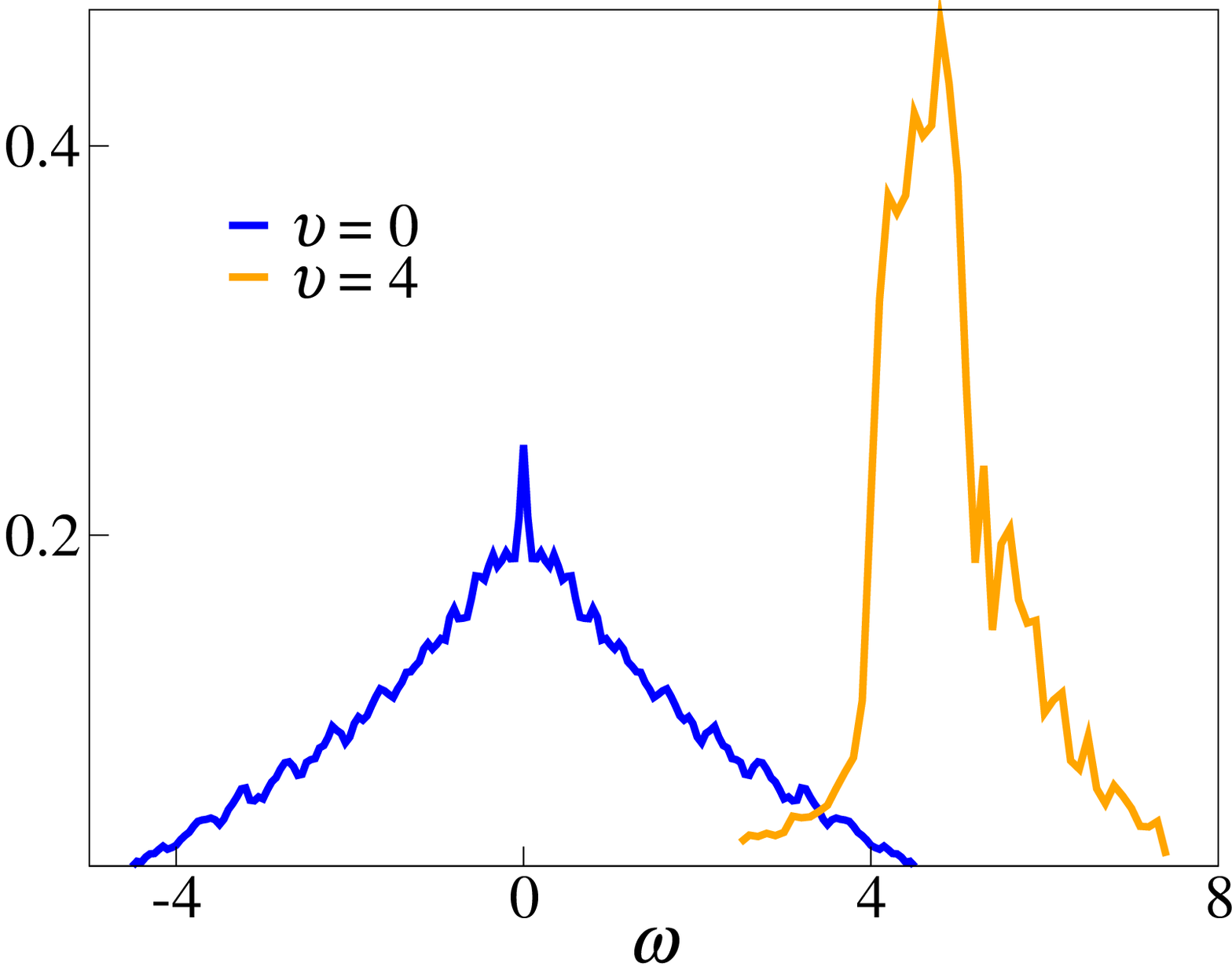}
\includegraphics[width = 0.328\textwidth]{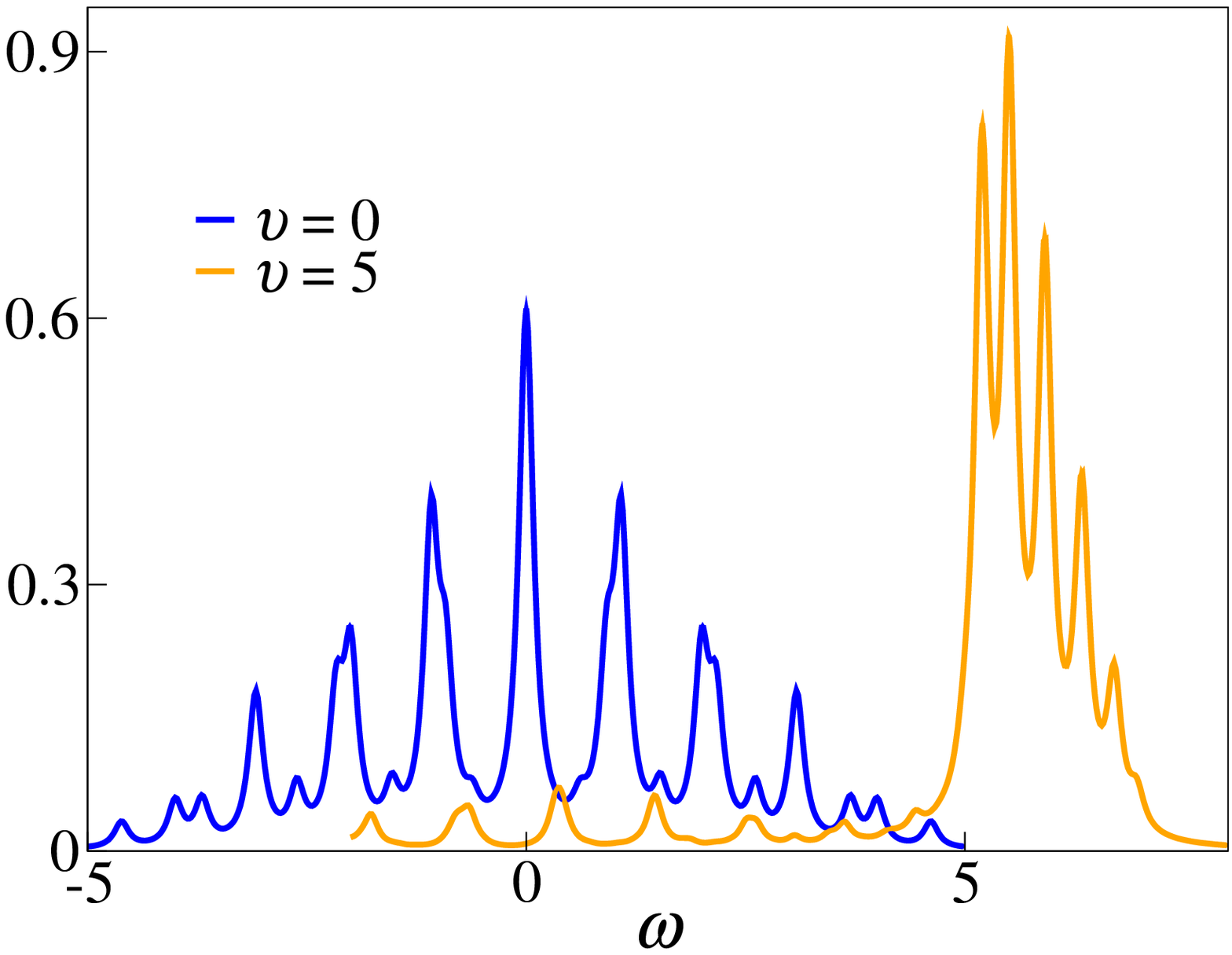}
\caption{Spectral functions of bound pairs at different nearest neighbour interaction strengths in 1D lattice ($L = 1000$) -left panel and 2D lattices (L = 80) - middle panel. c) Bound pair functions for onsite interaction of two particles in a Bethe lattice (L = 26) - right panel.}
\label{ideal}
\end{figure}

\begin{figure}[h]
\includegraphics[width = 0.328\textwidth]{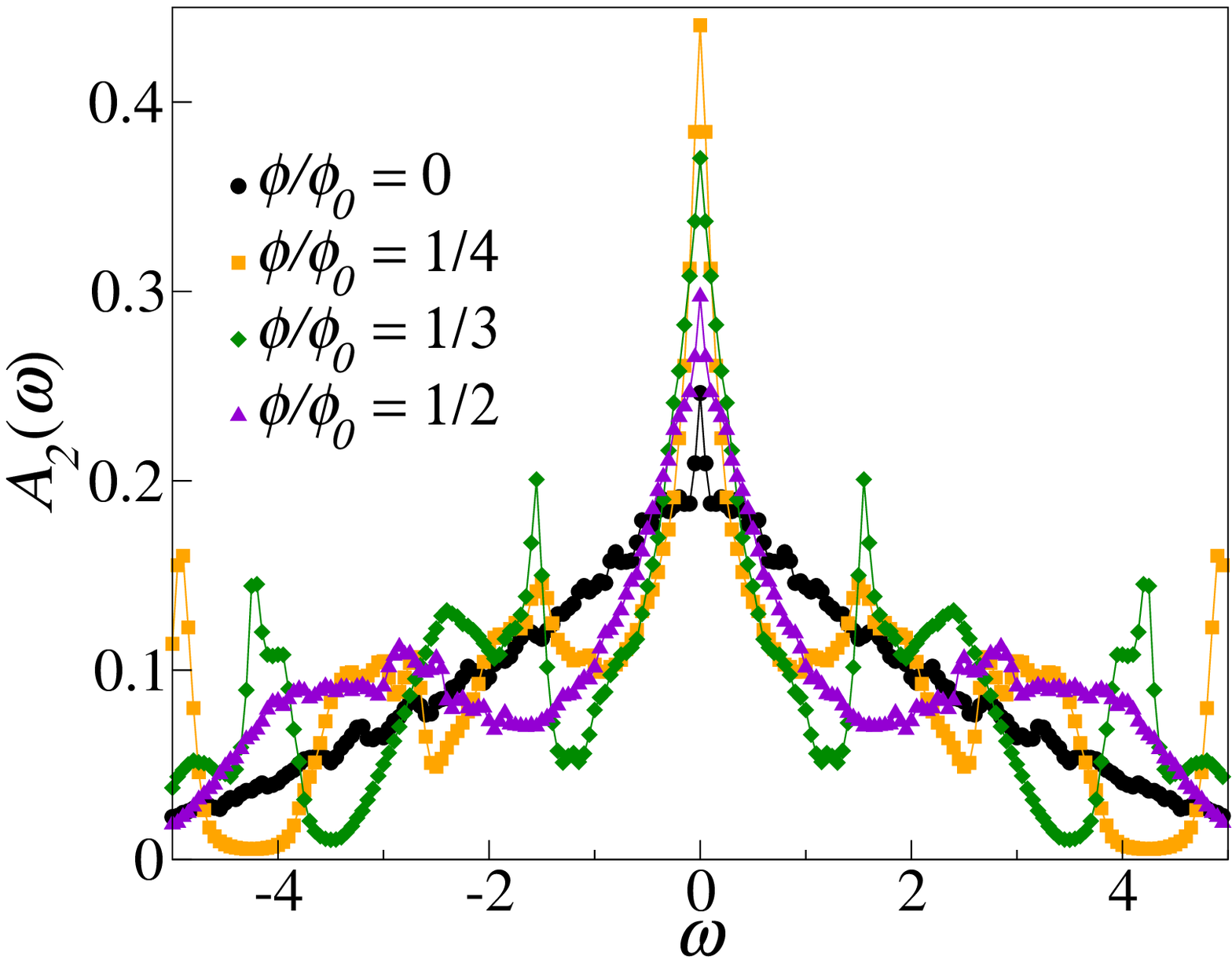}
\includegraphics[width = 0.328\textwidth]{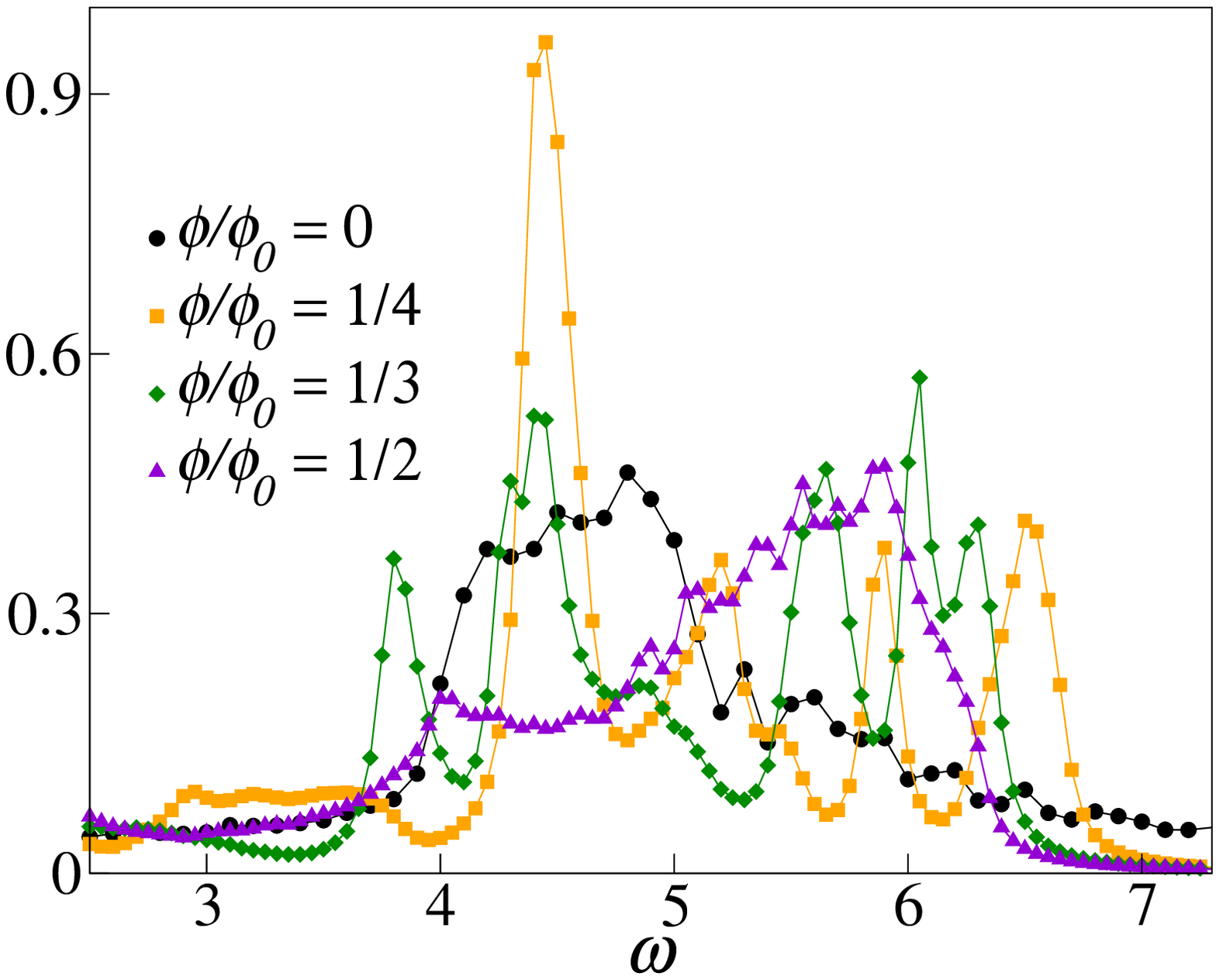}
\includegraphics[width = 0.328\textwidth]{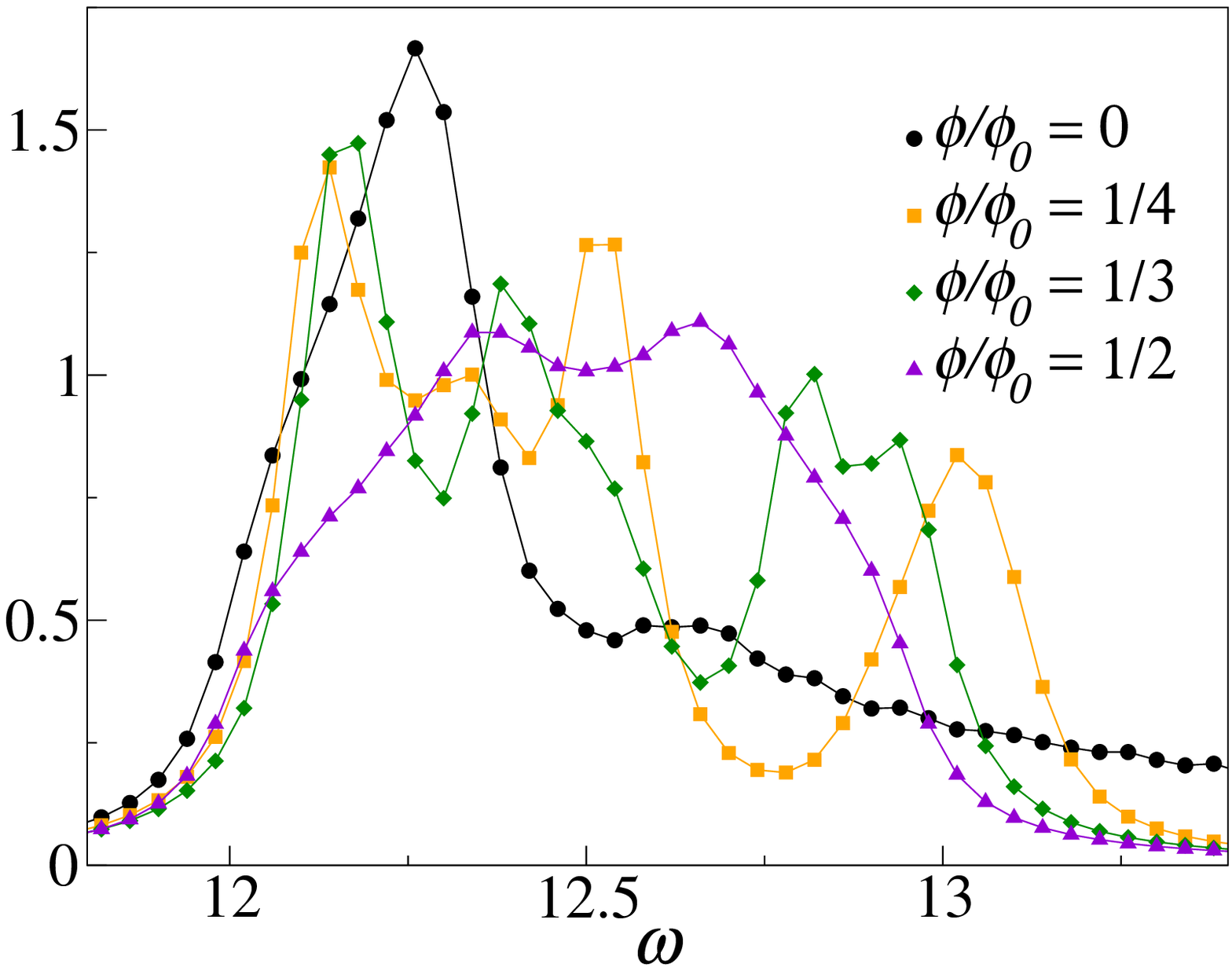}
\caption{Pair functions on a 2D lattice of 20 sites per dimension with magnetic flux per plaquette for $v_1=0$ - left panel, $v_1=4$ - middle panel, $v_1=12$ - right panel.}
\label{2D-external}
\end{figure}

\end{widetext}

The presence of flux per plaquette changes the spectral profiles significantly. For non-interacting particles the pair function shows multiple peaks depending on the flux fractions. In Fig. \ref{2D-external}, for simple fractions such as $\frac{1}{q}$, $q$ broad peaks are observed on each side of the fermi level ($\omega = 0$). For the interacting cases, generally $q$ broad peaks are observed. However these broad peaks contain multiple peaks of smaller width within them. The pair function at half flux with the fraction $\frac{1}{2}$ is very similar to Kondo spectral profile. These spectra with flux per plaquette can have flux origins from magnetic impurities within the lattice or from external magnetic fields. The spectra changes significantly with any minute changes with flux. This shows a better control over the pair properties may be achieved through quantized impurity spin fluxes within the lattice rather than with external control of the field. Although spectral widths remain same with flux changes, enhancement of few peaks with increasing $q$ (and more splitting) is observed which can enhance pair density within a range of $\omega$ with addition or removal of populated states from the lattice.

\section{Conclusion}
In this article the pair functions are obtained with a recursion algorithm that computes few-particle Green's functions efficiently in ordered and disordered lattices. The pair functions reveal dependence of pair population on interaction strengths in lattices with and without flux. The form of the discussed recursion is applicable beyond tridiagonalization which is the case for other efficient algorithms limited to single particle in 1D and beyond regular 1D, 2D and 3D lattices to graph structures as Bethe lattices. For disordered systems, the real space formulation allows for application of approximations that enhance efficiency without significant cost to accuracy. Although presented calculations here are limited to sites with single level systems without internal structures, similar calculations can be performed for lattice sites with internal structures within reasonable system sizes.

\newpage

\nocite{*}


\begin{thebibliography}{50}
%
%

\bibitem{berciu}
M. Berciu and A. M. Cook Efficient computation of lattice green’s functions for models with nearest-neighbour hopping. \textit{Eur. Phys. Lett.} \textbf{92} 40003 (2010)

\bibitem{economou}
E. N. Economou Green's Functions in Quantum Physics. \textit{Springer} (2010)

\bibitem{recursion}
D. G. Pettifor and D. L. Weaire (Eds.) The Recursion Method and Its Applications. \textit{Springer-Verlag} (1987)

\bibitem{morita}
T. Morita Useful procedure for computing the lattice green’s function-square, tetragonal, and bcc lattices. \textit{J. Math. Phys.} \textbf{12} 1744 (1971)

\bibitem{haydock}
R. Haydock, V. Heine and M J Kelly Electronic structure based on the local atomic environment for tight-binding bands:II. \textit{J. Phys. C: Solid State Phys.} \textbf{8} 2591 (1975)

\bibitem{lanczos}
C. Lanczos An iteration method for the solution of the eigenvalue Problem of linear differential and integral operators. \textit{J. Res. Natl. Bur. Std.} \textbf{45} 255 (1950)



\bibitem{cooper}
J. Bardeen, L. N. Cooper  and J. R. Schrieffer Theory of Superconductivity. \textit{Phys. Rev.} \textbf{108} 1175 (1957)

\bibitem{winkler}
K. Winkler, G. Thalhammer, F. Lang, R. Grimm, J. Hecker Denschlag, A. J. Daley, A. Kantian, H. P. Buchler and P. Zoller Repulsively bound atom pairs in an optical lattice. \textit{Nature} \textbf{441} 853 (2006)

\bibitem{greiner}
P. M. Preiss, R. Ma, M-E. Tai, A. Lukin, M. Rispoli, P. Zupancic, Y. Lahini, R. Islam and M. Greiner Strongly correlated quantum walks in optical lattices. \textit{Science} \textbf{347} 1229 (2015)

\bibitem{bloch}
T. Fukuhara, P. Schauß,  M. Endres, S. Hild, M. Cheneau, I. Bloch and C. Gross Microscopic observation of magnon bound states and their dynamics. \textit{Nature} \textbf{502} 76 (2013)

\bibitem{lahini1}
Y. Lahini, M. Verbin, S. D. Huber, Y. Bromberg, R. Pugatch and Y. Silberberg Quantum walk of two interacting bosons. \textit{Phys. Rev. A} \textbf{86} 011603(R)(2012)

\bibitem{tirtha1}
T. Chattaraj and R. V. Krems 2016 Effects of long-range hopping and interactions on quantum walks in ordered and disordered lattices. \textit{Phys. Rev. A} \textbf{94} 023601 (2016)

\bibitem{anderson}
P. W. Anderson The Theory of Superconductivity in the High-Tc Cuprates. \textit{Princeton Univ. Press} (1997)

\bibitem{rausch}
R. Rausch and M. Potthoff Multiplons in the two-hole excitation spectra of the one-dimensional Hubbard model. \textit{New J. Phys.} \textbf{18} 023033 (2016)

\bibitem{engel}
G. S. Engel, T. R. Calhoun, E. L. Read, T-K. Ahn, T. Mancal, Y-C. Cheng, R. E. Blankenship and G. R. Fleming Evidence for wavelike energy transfer through quantum coherence in photosynthetic systems. \textit{Nature} \textbf{446} 782 (2007)

\bibitem{sarovar}
M. Sarovar, A. Ishizaki, G. R. Fleming and K. B. Whaley Quantum entanglement in photosynthetic light-harvesting complexes. \textit{Nature Phys.} \textbf{6} 462 (2010)

\bibitem{schreiber}
R. A. Romer and M. Schreiber No enhancement of the localization length for two interacting particles in a random potential. \textit{Phys. Rev. Lett.} \textbf{78} 515 (1997)

\bibitem{pichard}
S. De Toro Arias, X. Waintal and J-L. Pichard Two interacting particles in a disordered chain III: Dynamical aspects of the interplay disorder-interaction. \textit{Eur. Phys. J. B} \textbf{10} 149 (1999)

\bibitem{flach}
D. O. Krimer, R. Khomeriki, and S. Flach Two interacting particles in a random potential. \textit{JETP Lett.} \textbf{94} 406 (2011)

\bibitem{tirtha2}
T. Chattaraj Localization Parameters for Two Interacting Particles in Disordered Two-Dimensional Finite Lattices. \textit{Condens. Matter} \textbf{3} 38 (2018)

\bibitem{ortuno}
M. Ortuno and E. Cuevas  Localized to extended states transition for two interacting particles in a two-dimensional random potential. \textit{Europhys. Lett.} \textbf{46} 224 (1999)

\bibitem{moeller}
M. Moeller, A. Mukherjee, C. P. J. Adolphs, D. J. J. Marchand and M. Berciu Efficient computation of lattice green functions for models with longer range hopping. \textit{J. Phys. A: Math. Theor.} \textbf{45} 115206 (2012)

\bibitem{lahini2}
Y. Lahini, Y. Bromberg, D. N. Christodoulides and Y. Silberberg Quantum correlations in two-particle anderson localization. \textit{Phys. Rev. Lett.} \textbf{105} 163905 (2010)

\bibitem{strohmaier}
N. Strohmaier, D. Greif, R. Jordens, L. Tarruell, H. Moritz, T. Esslinger, R. Sensarama, D. Pekker, E. Altman, and E. Demler Observation of Elastic Doublon Decay in the Fermi-Hubbard Model. \textit{Phys. Rev. Lett.} \textbf{104} 080401 (2010)

\end{thebibliography}

\end{document}